\documentclass[10pt,conference,compsocconf,final]{IEEEtran}
\usepackage{times}

\usepackage{caption}
\usepackage[noadjust]{cite}

\ifCLASSINFOpdf
\else
\fi

\usepackage{ulem}

\usepackage{color}
\usepackage{soul}

\begin{document}
\pagenumbering{gobble}
%
\title{Beyond the Dolev-Yao Model: \\ Realistic Application-Specific Attacker Models for Applications Using Vehicular Communication}

\author{%
\IEEEauthorblockN{Christoph Ponikwar, Hans-Joachim Hof}
\IEEEauthorblockA{
MuSe - Munich IT Security Research Group\\
Department of Computer Science and Mathematics\\Munich University of Applied Sciences (MUAS), Germany}
Email: {\tt christoph.ponikwar@hm.edu}, \\{\tt hof@hm.edu} %
\and
\IEEEauthorblockN{Smriti Gopinath, Lars Wischhof}
\IEEEauthorblockA{
~\\
Department of Computer Science and Mathematics\\Munich University of Applied Sciences (MUAS), Germany}
Email: {\tt smriti.gopinath@hm.edu}, \\{\tt wischhof@hm.edu} %
}


\maketitle

\begin{abstract}
In recent time, the standards for Vehicular Ad-hoc Networks (VANETs) and Intelligent Transportation Systems (ITSs) matured and scientific and industry interest is high especially as autonomous driving gets a lot of media attention. Autonomous driving and other assistance systems for cars make heavy use of VANETs to exchange information.They may provide more comfort, security and safety for drivers. However, it is of crucial importance for the user's trust in these assistance systems that they could not be influenced by malicious users. VANETs are likely attack vectors for such malicious users, hence application-specific security requirements must be considered during the design of applications using VANETs. In literature, many attacks on vehicular communication have been described but attacks on specific vehicular networking applications are often missing. This paper fills in this gap by describing standardized vehicular networking applications, defining and extending previous attacker models, and using the resulting new models to characterize the possible attackers interested in the specific vehicular networking application. The attacker models presented in this paper hopefully provide great benefit for the scientific community and industry as they allow to compare security evaluations of different works, characterize attackers, their intentions and help to plan application-specific security controls for vehicular networking applications. 
\end{abstract}



%
\IEEEpeerreviewmaketitle

\section{Introduction}
\label{sec:introduction}
Vehicular networking applications are a subset of applications used in Intelligent Transportation Systems (ITSs). They typically need security controls, especially, when safety is at stake. For a constructive planning of security controls, it is of benefit to have a model of a typical attacker, a so-called attacker model. Typical attack classes are impersonation, data tampering, sybil attacks, or Denial of Service (DOS) attacks, please refer to \cite{ponikwar2015overview} for a survey on these attacks. However, these attack classes are very general and their severity differs from application to application. Hence, it is beneficial to have application-specific attacker models for vehicular networking applications. This paper presents vehicular networking applications specific attacker models. These attacker models could be used for security control planning as well as evaluation of security controls. Also, standardized attacker models as in this paper are hopefully a great benefit for the scientific community to compare evaluations of different papers and modeling real world attackers.

This paper is structured as follows: Section \ref{related} presents related work and shows the gap this paper is closing. Section \ref{sec:vehicular-networking-applications} gives an overview on vehicular networking applications. Section \ref{sec:attacker-model} presents a classification of attackers that is used for the application-specific attacker models introduced in Section \ref{sec:app-specific-attacker-models}. Section \ref{sec:conclusion} concludes the paper.

\section{Related Work}
\label{related}
The field of attack modeling has a long history with some of it rooting in reliability engineering and the vault tree analysis which got adopted and adapted as attack trees \cite{salter1998toward,schneier1999attack,Moore2001} in the realm of secure systems engineering. Because of its detailed and explicit nature the attack tree modeling approach is best suited when goals of an attacker have been elicited and actual mitigation should be developed. The approach taken here categorizes attackers based on different aspects that are derived from their goal, which in return tries to take advantage of a specific vehicular networking application. Others use a game theory based approach to infer intentions, objectives and strategies of attackers \cite{Liu2005}, we derive these from the vehicular networking application that the attacker tries to exploit. 

The often cited Dolev-Yao attacker model \cite{Dolev1983} models the attacker as an active saboteur. He is omnipotent and can therefore intercept, eavesdrop, or modify all communication of the network. Furthermore, the attacker can pose as a legitimate communication partner and can therefore initiate a communication with every participant in the network. Compromising or breaking cryptographic primitives is not possible for a Dolev-Yao attacker. Networks in an Intelligent Transportation System (ITS) aren't limited to the Internet, instead they consist of Vehicular Ad-hoc Networks (VANET), enabling ad-hoc communication. Cellular technologies, like Long Term Evolution (LTE), can provide connectivity to the Internet. Roadside Units (RSU) or other stationary participants could be connected via traditional electrical or optical wired technologies to other separated networks or the Internet. The Dolev-Yao model is far too imprecise for such a complex networking structure and it only depicts a special type of attacker. This attacker is also unrealistically strong by being omnipotent, which gets increasingly unlikely the more complex and diverse a network becomes. This was previously pointed out in regards to sensor networks \cite{Hof2007}\cite{hofzitterbart2005}. Especially, it is pointed out that physical security should not be expected because an attacker can easily get access to those nodes and perform a take over or compromise cryptographic secrets \cite{Hof2007}. In such a way, an outside attacker becomes an inside participating one. To sum it up, the Dolev-Yao model is far too imprecise and unrealistically strong to be of use for security controls planning in realistic vehicular networking scenarios.

A realistic attack scenario is the exploitation of low level software or hardware vulnerabilities in the network stacks of wireless transceivers. The existence and importance of these vulnerabilities has been discussed in various publications, \cite{Mulliner2011,Mulliner2012,Weinmann2012,Weinmann2013}. This scenario marks the lower bound of attack scenarios that are discussed in this paper. While still being relevant specifically to wireless communication, cellular or ad-hoc, it is also not specific to only one vehicular networking application and the root cause of vulnerable soft- and hardware proliferates through all the layers of current systems and is not specific to wireless communications. Therefore, this is not in focus for this publication. Instead, the main contribution is the combination and extension of previous attacker models by \cite{Dolev1983}\cite{Hof2007}\cite{Raya2007a} and the detailed description of realistic attacker models via the extended model. Most of the previous works \cite{Amoozadeh2015,hoa_la_security_2014,Nikaein2012,Sumra2011,leinmuller2008modeling}, are missing realistic attacker models. Some like \cite{hoa_la_security_2014,Nikaein2012,Sumra2011} use categories of attacks, like impersonation, data tampering, sybil, or DOS attacks and describe each attacker based on its category. \cite{leinmuller2008modeling} is really close to defining realistic attacker models by defining categories of attackers, like driver, road side or infrastructure. 

Realistic attacker models are needed to better understand who might be the attacker of a system, for better comparison and ultimately needed to make risk based decisions about whether to implement security controls and how to guard against a specific realistic attacker. 

\section{Vehicular Networking Applications}
\label{sec:vehicular-networking-applications}
A general classification of vehicular networking applications uses two classes: \textit{safety applications} and \textit{non-safety applications}. For realistic attacker models, a more fine-grain categorization is needed. The classes used in this paper are described in the following. Please refer to \cite{ponikwar2015overview} for a detailed description of the vehicular networking applications. 

\subsection{Cooperative Sensing (Safety)} 
Cooperative Sensing applications use V2X communication for situation awareness, e.g., to reduce risks of accidents while driving.

  \textbf{Road Hazard Signalling (RHS):}  When a vehicle picks up a standardized condition \cite{european_telecommunications_standards_institute_etsi_2013}, an application broadcasts these conditions to other recipients using a Decentralized Environmental Notification Messages (DENM) \cite{european_telecommunications_standards_institute_etsi_2010-1}. Conditions include emergency vehicle approaching, slow vehicle, stationary vehicle,  emergency electronic brake lights, wrong way driving, adverse weather condition, hazardous location, traffic condition, roadwork, and human presence on the road. 

\textbf{Cooperative Collision Avoidance (CCA):} When a vehicle senses a possible collision with an approaching vehicle based on Cooperative Awareness Messages (CAM) \cite{european_telecommunications_standards_institute_etsi_2011} received from nearby vehicles, the driver gets a warning. Two distinct collision warning applications has been specified:  Intersection Collision Risk Warning (ICRW) (a warning is triggered if a collision is likely to happen at an intersection) and Longitudinal Collision Risk Warning (LCRW) ( a warning is displayed to the driver if a front or rear end collision is likely)\cite{european_telecommunications_standards_institute_etsi_2013-11}. 



\subsection{Cooperative Maneuvering} 
Applications apply V2X communication for driving automation functions in the levels 3 to 5 as defined in SAE J3016 \cite{sae_international_-_on-road_automated_vehicle_standards_committee_taxonomy_2014}.

\textbf{Cooperative Adaptive Cruise Control (CACC):} To optimize resource usage by forming a convoy or platooning and reducing speed alteration via an extended horizon where minor changes can be leveled out.

\textbf{Cooperative Merging Assistance (CMA):} To avoid collisions vehicles and roadside units (RSU) cooperate and  negotiate merging maneuvers.

\textbf{Cooperative Automated Overtake (CAO):} For takeover maneuvers either in a fully autonomous self-driving or a driver assistance scenario, cooperation among vehicles to improve safety is needed.

\subsection{In-Vehicle Internet Access} 
Internet-based applications are offered to passengers and in distraction reduced versions even to the driver.

\subsection{Mobility Monitoring and Configuration} 
The status of a vehicle can be remotely queried and modified. This application includes control of auxiliary heating systems as well as software and firmware updates. Usually, the accessed vehicle is in a parked position during the interactions of this application.

\section{Attacker Model}
\label{sec:attacker-model}
There are already different characteristics for attackers known in literature, some described in the following paragraphs and extended if needed. 

\textbf{Insider Attacker vs. Outsider Attacker} \cite{Hof2007}\cite{Raya2007a}: An outsider attacker is restricted because he does not participate in regular communication. An insider attacker on the other hand is a regular participant in the communication. A participant could become an insider attacker e.g., when hacked or infected with malware.

\textbf{Active Attacker vs. Passive Attacker} \cite{Hof2007}\cite{Raya2007a}: A passive attacker only eavesdrops on communication. An active attacker on the other hand acts in the network, e.g., by creating and inserting messages, by replaying messages, or by modifying existing messages.

\textbf{Static Attacker vs. Dynamic Attacker} \cite{Hof2007}: An attacker adapting his behavior based on the behavior of network environment or attack target is called a dynamic attacker. Static attackers on the other hand do not adapt to changes whatsoever. An example of a static attack is the most basic form of malware which doesn't utilize a command and control infrastructure and is build only for a specific purpose, like sending spam. An example of a dynamic attack is an attacker of an Advanced Persistent Threat campaign, which adapts to security measures or changes his goal based on the detected environment around it.

\textbf{Cooperative Attacker} \cite{Hof2007} \textbf{vs. Individual Attacker}: Attackers colluding to reach a common goal (e.g., destabilization of the network) are called cooperative attackers. An attacker limited to its own abilities is called an individual attacker.

 \textbf{Local Attacker vs. Extended Attacker} \cite{Raya2007a} \textbf{extension: Global Attacker} \cite{Hof2007}: How much influence an attacker has is an important criteria for the scope and impact a given attack can develop. Limited by his physical abilities, a local attacker can only influence participants in his ad-hoc communication vicinity.  An attacker controlling multiple network segments has the ability to execute more sophisticated attacks that need a greater area of influence. This so-called global attacker has the ability to access every message of the network. But based on the diversity and complexity of ITS network architecture, this type of attacker is limited to the infrastructure providers or to attackers that can influence or execute control over this communication infrastructure.

\textbf{Malicious Attacker vs. Rational Attacker} \cite{Hof2007}\cite{Raya2007a} \textbf{extension: Opportunistic Attacker}: An indiscriminate attacker who does not care about losses, resource usage, or consequences and targets functionality of participants or the network is called malicious attacker.  A rational attacker tries to reach a certain goal by the cheapest means possible and is focused on his benefit or profit. An attacker who only executes an attack when an opportune circumstance occurs is called an opportunistic attacker.

Table \ref{tab:attacker-properties-matrix} shows the profile matrix based on the attacker characteristics described above  that is used in the rest of this paper to describe application-specific attackers.

\captionsetup{font={footnotesize,sc},justification=centering,labelsep=period}%
\begin{table}[htbp]
\centering
\renewcommand{\arraystretch}{1.2}
\caption{General attacker profile matrix}  
\begin{tabular}{|l|c|c|c|c|}
\hline
\multicolumn{5}{|c|}{Attacker Properties} \\ \hline
Membership:     & \multicolumn{2}{c|}{insider}     & \multicolumn{2}{c|}{outsider}      \\ \hline
Method:         & \multicolumn{2}{c|}{active}      & \multicolumn{2}{c|}{passive}       \\ \hline
Adaptability:   & \multicolumn{2}{c|}{dynamic}     & \multicolumn{2}{c|}{static}        \\ \hline
Organization:   & \multicolumn{2}{c|}{cooperative} & \multicolumn{2}{c|}{individual}    \\ \hline
Scope:          &            global & extended     & \multicolumn{2}{c|}{local}         \\ \hline
Motivation:     &         malicious & rational     & \multicolumn{2}{c|}{opportunistic} \\ \hline
\end{tabular}
\label{tab:attacker-properties-matrix} 
\end{table}

Based on this profile matrix, specific attackers can be modeled. The worst possible attacker is shown in Table \ref{tab:worst-attacker-properties-matrix}. The worst possible attacker is the most powerful attacker one can think of.  As described in Section \ref{related} for the Dolev-Yao model, such a powerful attacker is quite unlikely to appear in most realistic scenarios (however, there is one valid scenario listed below).


\captionsetup{font={footnotesize,sc},justification=centering,labelsep=period}%
\begin{table}[htbp]
\centering
\renewcommand{\arraystretch}{1.2}
\caption{Worst attacker profile matrix}  
\begin{tabular}{|l|c|c|c|c|}
\hline
\multicolumn{5}{|c|}{Attacker Properties} \\ \hline
Membership:     & \multicolumn{2}{c|}{insider}        & \multicolumn{2}{c|}{\sout{outsider}}      \\ \hline
Method:         & \multicolumn{2}{c|}{active}         & \multicolumn{2}{c|}{\sout{passive}}       \\ \hline
Adaptability:   & \multicolumn{2}{c|}{dynamic}        & \multicolumn{2}{c|}{\sout{static}}        \\ \hline
Organization:   & \multicolumn{2}{c|}{cooperative}    & \multicolumn{2}{c|}{\sout{individual}}    \\ \hline
Scope:          &            global & \sout{extended} & \multicolumn{2}{c|}{\sout{local}}         \\ \hline
Motivation:     &         malicious & \sout{rational} & \multicolumn{2}{c|}{\sout{opportunistic}} \\ \hline
\end{tabular}
\label{tab:worst-attacker-properties-matrix} 
\end{table}

Table \ref{tab:weakest-attacker-properties-matrix} shows the weakest possible attacker of the application specific attacker model. 


\captionsetup{font={footnotesize,sc},justification=centering,labelsep=period}%
\begin{table}[htbp]
\centering
\renewcommand{\arraystretch}{1.2}
\caption{Weakest attacker profile matrix}  
\begin{tabular}{|l|c|c|c|c|}
\hline
\multicolumn{5}{|c|}{Attacker Properties} \\ \hline
Membership:     & \multicolumn{2}{c|}{\sout{insider}}     & \multicolumn{2}{c|}{outsider}      \\ \hline
Method:         & \multicolumn{2}{c|}{\sout{active}}      & \multicolumn{2}{c|}{passive}       \\ \hline
Adaptability:   & \multicolumn{2}{c|}{\sout{dynamic}}     & \multicolumn{2}{c|}{static}        \\ \hline
Organization:   & \multicolumn{2}{c|}{\sout{cooperative}} & \multicolumn{2}{c|}{individual}    \\ \hline
Scope:          &     \sout{global} & \sout{extended}     & \multicolumn{2}{c|}{local}         \\ \hline
Motivation:     &  \sout{malicious} & \sout{rational}     & \multicolumn{2}{c|}{opportunistic} \\ \hline
\end{tabular}
\label{tab:weakest-attacker-properties-matrix} 
\end{table}

The worst attacker and weakest attacker are both ends of the application specific attacker model presented in this paper. However, in most vehicular networking applications a realistic attacker model lies between the worst attacker and the weakest attacker. The following section presents the realistic attacker models applicable for each vehicular networking application presented in Section \ref{sec:vehicular-networking-applications}.

\section{Application Specific Attacker Models}
\label{sec:app-specific-attacker-models}
For each vehicular networking application (see Section \ref{sec:vehicular-networking-applications}.), different specific attacker profiles are described in this section. 

\subsection{Cooperative Sensing (Safety)}
\label{subsec:cooperative-sensing-safety}
Attackers interfering with safety functions are always inadvertently or intentionally risking to cause damage to themselves or other humans besides causing financial damage. It is important to keep this in mind especially when judging about the motivation of a certain attacker.

A perpetrator is stuck in traffic, he then decides to push a button that forces his vehicle to send out false road hazard warnings to influence other vehicles. In an ideal situation for the attacker, the victim vehicles fall for his false claims. He might pose as an emergency vehicle, send out false wrong way driving warnings, roadwork, or human presence on the road to clear a lane, to speed past other vehicles. He is an active dynamic insider acting as an individual, with local reach, see table \ref{tab:speedster-properties-matrix}. As stated previously, fiddling with safety functions is borderline malicious activity. The speeding attacker might still try to be rational about the reliance of the successful deceiving of other traffic participants as they might simply ignore his false claims or he might overlook real hazards. 

\captionsetup{font={footnotesize,sc},justification=centering,labelsep=period}%
\begin{table}[htbp]
\centering
\renewcommand{\arraystretch}{1.2}
\caption{Speedster profile matrix}  
\begin{tabular}{|l|c|c|c|c|}
\hline
\multicolumn{5}{|c|}{Attacker Properties} \\ \hline
Membership:     & \multicolumn{2}{c|}{insider}               & \multicolumn{2}{c|}{\sout{outsider}}      \\ \hline
Method:         & \multicolumn{2}{c|}{active}                & \multicolumn{2}{c|}{\sout{passive}}       \\ \hline
Adaptability:   & \multicolumn{2}{c|}{dynamic}               & \multicolumn{2}{c|}{\sout{static}}        \\ \hline
Organization:   & \multicolumn{2}{c|}{\sout{cooperative}}    & \multicolumn{2}{c|}{individual}           \\ \hline
Scope:          &            \sout{global} & \sout{extended} & \multicolumn{2}{c|}{local}                \\ \hline
Motivation:     &         \sout{malicious} & rational        & \multicolumn{2}{c|}{\sout{opportunistic}} \\ \hline
\end{tabular}
\label{tab:speedster-properties-matrix} 
\end{table}

Another group taking advantage of this safety application may be a single or group of environmentalists or annoyed residents. Their goal might be to reduce the speed of vehicles, no matter what the rest of the community decided on to be acceptable. There are two basic technical approaches these attacker can pursue either they try to jam valid RSUs (Denial of Service), see Table \ref{tab:outsider-traffic-calming-properties-matrix}, or they try to compromise or mimic a valid RSU, see Table\ref{tab:insider-traffic-calming-properties-matrix}.

\captionsetup{font={footnotesize,sc},justification=centering,labelsep=period}%
\begin{table}[htbp]
\centering
\renewcommand{\arraystretch}{1.2}
\caption{Outsider Traffic Calming profile matrix}  
\begin{tabular}{|l|c|c|c|c|}
\hline
\multicolumn{5}{|c|}{Attacker Properties} \\ \hline
Membership:     & \multicolumn{2}{c|}{\sout{insider}}        & \multicolumn{2}{c|}{outsider}      \\ \hline
Method:         & \multicolumn{2}{c|}{active}                & \multicolumn{2}{c|}{\sout{passive}}\\ \hline
Adaptability:   & \multicolumn{2}{c|}{\sout{dynamic}}        & \multicolumn{2}{c|}{static}        \\ \hline
Organization:   & \multicolumn{2}{c|}{\sout{cooperative}}    & \multicolumn{2}{c|}{individual}    \\ \hline
Scope:          &            \sout{global} & \sout{extended} & \multicolumn{2}{c|}{local}         \\ \hline
Motivation:     &         \sout{malicious} & \sout{rational} & \multicolumn{2}{c|}{opportunistic} \\ \hline
\end{tabular}
\label{tab:outsider-traffic-calming-properties-matrix} 
\end{table}

When being able to communicate to other vehicles other attacks are possible, like trying to get the vehicles to alter their route, because of hazard warnings like weather conditions or fake traffic conditions. But a single RSU or a fake one has only a limited area of influence.

\captionsetup{font={footnotesize,sc},justification=centering,labelsep=period}%
\begin{table}[htbp]
\centering
\renewcommand{\arraystretch}{1.2}
\caption{Insider Traffic Calming profile matrix}  
\begin{tabular}{|l|c|c|c|c|}
\hline
\multicolumn{5}{|c|}{Attacker Properties} \\ \hline
Membership:     & \multicolumn{2}{c|}{insider}               & \multicolumn{2}{c|}{\sout{outsider}} \\ \hline
Method:         & \multicolumn{2}{c|}{active}                & \multicolumn{2}{c|}{\sout{passive}}  \\ \hline
Adaptability:   & \multicolumn{2}{c|}{\sout{dynamic}}        & \multicolumn{2}{c|}{static}          \\ \hline
Organization:   & \multicolumn{2}{c|}{\sout{cooperative}}    & \multicolumn{2}{c|}{individual}      \\ \hline
Scope:          &            \sout{global} & \sout{extended} & \multicolumn{2}{c|}{local}           \\ \hline
Motivation:     &         \sout{malicious} & \sout{rational} & \multicolumn{2}{c|}{opportunistic}   \\ \hline
\end{tabular}
\label{tab:insider-traffic-calming-properties-matrix} 
\end{table}

A small step up for the attacker who is compromising RSUs, see Table \ref{tab:insider-traffic-calming-properties-matrix}, to slow vehicles down, would be if he does not stop after controlling one RSU. He would try to get control over a larger area to have a bigger influence on victim vehicles, see Table \ref{tab:sophisticated-traffic-manipulation-properties-matrix}. By doing so he poses a greater risk to safety in that area by exercising his power over an area and colluding RSUs, to make the false or modified warnings look authentic. 

\captionsetup{font={footnotesize,sc},justification=centering,labelsep=period}%
\begin{table}[htbp]
\centering
\renewcommand{\arraystretch}{1.2}
\caption{Sophisticated Traffic Manipulation profile matrix}  
\begin{tabular}{|l|c|c|c|c|}
\hline
\multicolumn{5}{|c|}{Attacker Properties} \\ \hline
Membership:     & \multicolumn{2}{c|}{insider}        & \multicolumn{2}{c|}{\sout{outsider}}      \\ \hline
Method:         & \multicolumn{2}{c|}{active}         & \multicolumn{2}{c|}{\sout{passive}}       \\ \hline
Adaptability:   & \multicolumn{2}{c|}{dynamic}        & \multicolumn{2}{c|}{\sout{static}}        \\ \hline
Organization:   & \multicolumn{2}{c|}{cooperative}    & \multicolumn{2}{c|}{\sout{individual}}    \\ \hline
Scope:          &            \sout{global} & extended & \multicolumn{2}{c|}{\sout{local}}         \\ \hline
Motivation:     &         \sout{malicious} & rational & \multicolumn{2}{c|}{\sout{opportunistic}} \\ \hline
\end{tabular}
\label{tab:sophisticated-traffic-manipulation-properties-matrix} 
\end{table}

To prevent duplicate information, the following attack model omits the table, because the attacker resembles the worst case attacker as pictured in Table \ref{tab:worst-attacker-properties-matrix}. The attacker could be a foreign power either state sponsored or independent but the goal of this group would be to put on a stranglehold on safety related functions to unleash massive chaos by using infrastructure to flood victims with false hazard, collision warning and creating non existent vulnerable road users in front of vehicle, to get the safety systems to collapse and shutdown. During such an attack the goal of the attacker would be to create human casualties or at least create huge financial losses and impediment. The whole purpose of such a malicious attack is to weaken the position of an opponent and to strengthen their own, this could be also achieved by holding the infrastructure ransom and threatening to vandalize the infrastructure. To have such a large scale effect the attacker needs to compromise the infrastructure by ether common vulnerabilities or by compromising the provider of it.

The last three attacker types in this section dedicated are derived from the weakest attacker, see Table \ref{tab:weakest-attacker-properties-matrix}. The goal of these attackers is to acquire knowledge about nearby vehicles. This goal is similar to the snooping individual who uses the manufacturer build in monitoring as described in Table \ref{tab:control-freak-properties-matrix}. The difference between these three type of attackers is their scope, whether they have only local, extended or global reception. A local influence might be easy to establish, only one receiver is needed. For extended visibility, more receivers are required, but for global reach the RSU to attacker receiver ratio must be one-to-one. This would be easy to achieve for an worst case attacker as he does not only want to have control over some infrastructure but wants to have control over all available ones. 

\subsection{Cooperative Sensing (Information/Non-Safety)}
\label{subsec:cooperative-sensing-non-safety}
In comparison to the safety relevant applications mentioned before, informational cooperative sensing application do not have an immediate life threatening aspect. The application for exchanging dynamic mapping information is particularly interesting as it might be used to improve the driver's experience, but could be misused to annoy the driver or even to literally navigate him into dangerous situations. One attacker who is trying to annoy drivers or shop owners sets up a fake RSU to send out false information about points of interest. This might reach from false opening hours to false location information. This can be considered as trolling, wasting someone's time and resources and annoying people to no end, as presented in Table \ref{tab:trolling-false-information-attacker-properties-matrix}. He is rather static, a individual opportunistic attacker with only a local scope.

\captionsetup{font={footnotesize,sc},justification=centering,labelsep=period}%
\begin{table}[htbp]
\centering
\renewcommand{\arraystretch}{1.2}
\caption{Trolling via false information profile matrix}  
\begin{tabular}{|l|c|c|c|c|}
\hline
\multicolumn{5}{|c|}{Attacker Properties} \\ \hline
Membership:     & \multicolumn{2}{c|}{insider}               & \multicolumn{2}{c|}{\sout{outsider}} \\ \hline
Method:         & \multicolumn{2}{c|}{active}                & \multicolumn{2}{c|}{\sout{passive}}  \\ \hline
Adaptability:   & \multicolumn{2}{c|}{\sout{dynamic}}        & \multicolumn{2}{c|}{static}          \\ \hline
Organization:   & \multicolumn{2}{c|}{\sout{cooperative}}    & \multicolumn{2}{c|}{individual}      \\ \hline
Scope:          &            \sout{global} & \sout{extended} & \multicolumn{2}{c|}{local}           \\ \hline
Motivation:     &         \sout{malicious} & \sout{rational} & \multicolumn{2}{c|}{opportunistic}   \\ \hline
\end{tabular}
\label{tab:trolling-false-information-attacker-properties-matrix} 
\end{table}

The second type of attacker are criminals, see Table \ref{tab:criminal-false-information-attacker-properties-matrix}, that use technology to make their activities easier. In case of mapping information, they could try to trick the driver via the navigation system to take another route, to send the driver to an abundant place to either rob or kidnap him. It may be enough to setup some fake RSUs or compromise a few, software wise or physically, to mislead or manipulate the victims systems. A single criminal or a group of them may feed dynamic false information into the systems near their victim and may even deploy multiple RSUs to have a higher chance of misleading the driver. When considering criminals as attackers, the differentiation whether their motivation is malicious or rational depends on where the perpetrators want to reuse their scheme, like a business, or if they are outright hitmen. But whether the latter one would invest in the technology and know-how to ease his job of executing a paid for assassination is questionable. Nonetheless intentional criminal activity would be considered malicious. 

\captionsetup{font={footnotesize,sc},justification=centering,labelsep=period}%
\begin{table}[htbp]
\centering
\renewcommand{\arraystretch}{1.2}
\caption{False information as criminal activity support}  
\begin{tabular}{|l|c|c|c|c|}
\hline
\multicolumn{5}{|c|}{Attacker Properties} \\ \hline
Membership:     & \multicolumn{2}{c|}{insider}        & \multicolumn{2}{c|}{\sout{outsider}}        \\ \hline
Method:         & \multicolumn{2}{c|}{active}         & \multicolumn{2}{c|}{\sout{passive}}         \\ \hline
Adaptability:   & \multicolumn{2}{c|}{dynamic}        & \multicolumn{2}{c|}{\sout{static}}          \\ \hline
Organization:   & \multicolumn{2}{c|}{cooperative}    & \multicolumn{2}{c|}{\sout{individual}}      \\ \hline
Scope:          &            \sout{global} & extended & \multicolumn{2}{c|}{\sout{local}}           \\ \hline
Motivation:     &         malicious & \sout{rational} & \multicolumn{2}{c|}{\sout{opportunistic}}   \\ \hline
\end{tabular}
\label{tab:criminal-false-information-attacker-properties-matrix} 
\end{table}

\subsection{Cooperative Maneuvering}
\label{subsec:cooperative-maneuvering}
When considering cooperative maneuvering, one distinguishes if a non-cooperative fallback is available or not. If a non-cooperative fallback is available, the attacker might be just like the trolling one mentioned in Table \ref{tab:trolling-false-information-attacker-properties-matrix} as no real harm is possible because a safe downgrade to non cooperative assistance is available. If no fallback is available, there is a safety issue. CACC should have a non cooperative companion ACC. For the cooperative automated overtake application especially in an autonomous driving environment, the safety implications are obvious. An attacker sending false awareness information is only different from the worst case attacker (see Table \ref{tab:worst-attacker-properties-matrix}) in regards to his reach as he is locally limited and to the organizational aspect as he is an individual, see Table \ref{tab:fake-awareness-attacker-properties-matrix}.

\captionsetup{font={footnotesize,sc},justification=centering,labelsep=period}%
\begin{table}[htbp]
\centering
\renewcommand{\arraystretch}{1.2}
\caption{Intentionally false CAM attacker profile matrix}  
\begin{tabular}{|l|c|c|c|c|}
\hline
\multicolumn{5}{|c|}{Attacker Properties} \\ \hline
Membership:     & \multicolumn{2}{c|}{insider}               & \multicolumn{2}{c|}{\sout{outsider}}      \\ \hline
Method:         & \multicolumn{2}{c|}{active}                & \multicolumn{2}{c|}{\sout{passive}}       \\ \hline
Adaptability:   & \multicolumn{2}{c|}{dynamic}               & \multicolumn{2}{c|}{\sout{static}}        \\ \hline
Organization:   & \multicolumn{2}{c|}{\sout{cooperative}}    & \multicolumn{2}{c|}{individual}           \\ \hline
Scope:          &            \sout{global} & \sout{extended} & \multicolumn{2}{c|}{local}                \\ \hline
Motivation:     &                malicious & \sout{rational} & \multicolumn{2}{c|}{\sout{opportunistic}} \\ \hline
\end{tabular}
\label{tab:fake-awareness-attacker-properties-matrix} 
\end{table}

\subsection{In-Vehicle Internet Access}
\label{subsec:in-vehicular-internet-access}
A malware author who uses the Internet connectivity as an initial attack vector to infect software components in a vehicle is summarized in Table \ref{tab:vehicular-malware-initial-vector-properties-matrix}. This type of an active attack depends heavily on the design of the vehicular internet access capabilities. If the vehicle itself does not have Internet enabled or capable components and merely provides an access point for other smart devices to get access, than the attack surface is reduced. Still, an outside attacker could try to attack the access point software or more generally common software components among vehicles of the same manufacturer or across the industry, that is reachable via the Internet. The ability of an attacker to adapt his malware or the ability of it getting new orders via an command and control infrastructure makes him an dynamic opponent. As an individual attacker who uses the Internet as the initial access vector to his victims, his capabilities are also limited by the ability to directly connect to a victim or whether the victim has to make the initial connection. In this case, he would resort to common scenarios like water hole, or phishing attacks, where the victim connects to an Internet resource who serves an exploit kit targeted at software vulnerabilities. Nevertheless the attackers scope is limited in the sense of the initial attack vector to a local one, further more he is going to act in a rational way, as he wants to make a profit of off his work.

\captionsetup{font={footnotesize,sc},justification=centering,labelsep=period}%
\begin{table}[htbp]
\centering
\renewcommand{\arraystretch}{1.2}
\caption{Vehicular malware initial attack vector}  
\begin{tabular}{|l|c|c|c|c|}
\hline
\multicolumn{5}{|c|}{Attacker Properties} \\ \hline
Membership:     & \multicolumn{2}{c|}{\sout{insider}}        & \multicolumn{2}{c|}{outsider}             \\ \hline
Method:         & \multicolumn{2}{c|}{active}                & \multicolumn{2}{c|}{\sout{passive}}       \\ \hline
Adaptability:   & \multicolumn{2}{c|}{dynamic}               & \multicolumn{2}{c|}{\sout{static}}        \\ \hline
Organization:   & \multicolumn{2}{c|}{\sout{cooperative}}    & \multicolumn{2}{c|}{individual}           \\ \hline
Scope:          &            \sout{global} & \sout{extended} & \multicolumn{2}{c|}{local}                \\ \hline
Motivation:     &         \sout{malicious} & rational        & \multicolumn{2}{c|}{\sout{opportunistic}} \\ \hline
\end{tabular}
\label{tab:vehicular-malware-initial-vector-properties-matrix} 
\end{table} 

\subsection{Mobility Monitoring and Configuration}
\label{subsec:mobility-monitoring-and-configuration}
There are cases where an owner or an agent of the owner (modder, tuner) could be seen as an attacker from the perspective of a vehicle manufacturer. In this case, the owner or his agent tries to manipulate the vehicle, e.g., to decrease the mileage count of a car. It is obvious that the owner or his agent can access all available communication, hence he is an insider attacker. He also has the ability to modify the hardware of software and react to security controls in place. For example, extraction of cryptographic keys from firmware images is a well-known approach in the car hacking and chip tuning community. Hence, the attacker is an adaptive attacker. Attacks usually affect only one vehicle. A special case is an attack on an online service portal of the manufacturer. If all vehicles of this manufacturer can be modified remotely, the attack could have an extended scope, but the initial vulnerability is still local to the service portal. The owner of a vehicle is a rational attacker as he is resource sensitive. If the use of a vehicle hack has less value than the money needed to execute the hack, the owner likely will not execute the attack. See Table \ref{tab:modder-properties-matrix} for a summary.

\captionsetup{font={footnotesize,sc},justification=centering,labelsep=period}%
\begin{table}[htbp]
\centering
\renewcommand{\arraystretch}{1.2}
\caption{Modder/Tuner profile matrix}  
\begin{tabular}{|l|c|c|c|c|}
\hline
\multicolumn{5}{|c|}{Attacker Properties} \\ \hline
Membership:     & \multicolumn{2}{c|}{insider}             & \multicolumn{2}{c|}{\sout{outsider}}      \\ \hline
Method:         & \multicolumn{2}{c|}{active}              & \multicolumn{2}{c|}{\sout{passive}}       \\ \hline
Adaptability:   & \multicolumn{2}{c|}{dynamic}             & \multicolumn{2}{c|}{\sout{static}}        \\ \hline
Organization:   & \multicolumn{2}{c|}{\sout{cooperative}}  & \multicolumn{2}{c|}{individual}           \\ \hline
Scope:          &          \sout{global} & \sout{extended} & \multicolumn{2}{c|}{local}                \\ \hline
Motivation:     &       \sout{malicious} & rational        & \multicolumn{2}{c|}{\sout{opportunistic}} \\ \hline
\end{tabular}
\label{tab:modder-properties-matrix} 
\end{table}

Another attacker is a control freak attacker. His goal is snooping on his or her spouse, child, or anybody else using the vehicle.
As the owner of the vehicle, the active insider individual attacker can use the location tracking or monitoring ability for the legitimate purpose (e.g., finding his vehicle or creating an automatic driver's logbook) but also use it to spy on persons he lends the vehicle to. He does not need to change his behavior as tracking devices are already build into most vehicles. He is very opportunistic as he uses the abilities of the existing monitoring system. Only his own vehicle is affected. The properties of the control freak attacker are summarized in Table \ref{tab:control-freak-properties-matrix}.

\captionsetup{font={footnotesize,sc},justification=centering,labelsep=period}%
\begin{table}[htbp]
\centering
\renewcommand{\arraystretch}{1.2}
\caption{Control Freak profile matrix}  
\begin{tabular}{|l|c|c|c|c|}
\hline
\multicolumn{5}{|c|}{Attacker Properties} \\ \hline
Membership:     & \multicolumn{2}{c|}{insider}             & \multicolumn{2}{c|}{\sout{outsider}} \\ \hline
Method:         & \multicolumn{2}{c|}{active}              & \multicolumn{2}{c|}{\sout{passive}}  \\ \hline
Adaptability:   & \multicolumn{2}{c|}{\sout{dynamic}}      & \multicolumn{2}{c|}{static}          \\ \hline
Organization:   & \multicolumn{2}{c|}{\sout{cooperative}}  & \multicolumn{2}{c|}{individual}      \\ \hline
Scope:          &          \sout{global} & \sout{extended} & \multicolumn{2}{c|}{local}           \\ \hline
Motivation:     &       \sout{malicious} & \sout{rational} & \multicolumn{2}{c|}{opportunistic}   \\ \hline
\end{tabular}
\label{tab:control-freak-properties-matrix} 
\end{table}

An extension of the control freak attacker is an attacker attacking a centralized location information system of a manufacturer. If such a centralized system (e.g., a service portal) exists and the user can query it for the position of his vehicle (e.g., to find a parked car), it could be an attractive target. The attacker is an outside attacker but he must be highly motivated, persistent, and dynamic. When attacking the system, the possession or control of multiple vehicles might be advantageous but the attacker is still considered to be individual and locally limited to the attacked system, that stores the location information. The attacker is not interested to create outages or service interruption as he is interested in the functioning system and especially in the data it gathers, therefor he can be considered being rational. See Table \ref{tab:surveillance-attacker-properties-matrix} for a summary of this attacker.

\captionsetup{font={footnotesize,sc},justification=centering,labelsep=period}%
\begin{table}[htbp]
\centering
\renewcommand{\arraystretch}{1.2}
\caption{Mass Surveillance profile matrix}  
\begin{tabular}{|l|c|c|c|c|}
\hline
\multicolumn{5}{|c|}{Attacker Properties} \\ \hline
Membership:     & \multicolumn{2}{c|}{\sout{insider}}        & \multicolumn{2}{c|}{outsider}             \\ \hline
Method:         & \multicolumn{2}{c|}{active}                & \multicolumn{2}{c|}{\sout{passive}}       \\ \hline
Adaptability:   & \multicolumn{2}{c|}{dynamic}               & \multicolumn{2}{c|}{\sout{static}}        \\ \hline
Organization:   & \multicolumn{2}{c|}{\sout{cooperative}}    & \multicolumn{2}{c|}{individual}           \\ \hline
Scope:          &            \sout{global} & \sout{extended} & \multicolumn{2}{c|}{local}                \\ \hline
Motivation:     &         \sout{malicious} & rational        & \multicolumn{2}{c|}{\sout{opportunistic}} \\ \hline
\end{tabular}
\label{tab:surveillance-attacker-properties-matrix} 
\end{table}

The last two attacker models are still fit into the V2X communication and application paradigm, although they are centered around the existence of systems run by the manufacturer and misusing or exploiting weaknesses in them, which are reachable via the Internet.  

\section{Conclusion and Future Work}
\label{sec:conclusion}
This paper presented a survey on current vehicular networking applications, including Cooperative Sensing (Safety), Cooperative Sensing (Information/Non-Safety), Cooperative Maneuvering, In-Vehicle Internet Access, and Mobility Monitoring and Configuration. Novel attacker models are presented that focus on realistic application-specific attacks instead of general attacks on vehicular networks. 

\captionsetup{font={footnotesize,sc},justification=centering,labelsep=period}%
\begin{table}[htbp]
\centering
\renewcommand{\arraystretch}{1.2}
\caption{Attacker Model Overview}  
\begin{tabular}{|c|l|c|}
\hline
\multicolumn{3}{|c|}{Attacker Properties} \\ \hline
  1 & Speedster                                         & \ref{tab:speedster-properties-matrix} \\ \hline
  2 & Outsider Traffic Calming                          & \ref{tab:outsider-traffic-calming-properties-matrix} \\ \hline
  3 & Insider Traffic Calming                           & \ref{tab:insider-traffic-calming-properties-matrix} \\ \hline
  4 & Sophisticated Traffic Manipulation                & \ref{tab:sophisticated-traffic-manipulation-properties-matrix} \\ \hline
  5 & Massive Financial Damages and Human Casualties    & \ref{tab:worst-attacker-properties-matrix} \\ \hline
6-8 & Information Gathering with three different scopes & \ref{tab:weakest-attacker-properties-matrix} \\ \hline
  9 & Trolling via false information                    & \ref{tab:trolling-false-information-attacker-properties-matrix} \\ \hline
 10 & False information as criminal activity support    & \ref{tab:criminal-false-information-attacker-properties-matrix} \\ \hline
 11 & Intentionally false CAM attacker                  & \ref{tab:fake-awareness-attacker-properties-matrix} \\ \hline
 12 & Vehicular malware initial attack vector           & \ref{tab:vehicular-malware-initial-vector-properties-matrix}  \\ \hline
 13 & Modder/Tuner                                      & \ref{tab:modder-properties-matrix} \\ \hline
 14 & Control Freak                                     & \ref{tab:control-freak-properties-matrix} \\ \hline
 15 & Mass Surveillance                                 & \ref{tab:surveillance-attacker-properties-matrix} \\ \hline
\end{tabular}
\label{tab:overview-table-attacker-profiles} 
\end{table}

Our contribution describes 15 realistic attacker profiles in its main Section \ref{sec:app-specific-attacker-models}, an summary is given in table \ref{tab:overview-table-attacker-profiles}. These attacker models allow for a more focused planning of security controls for vehicular networks, as well as a better comparability of security evaluations using these attacker models.

Using this attacker modeling approach for evaluation and providing in-depth examples on how to benefit from it in particular vehicular communication applications is reserved for future work.



%
%
%

\bibliographystyle{IEEEtran}
\bibliography{securware2016}

\end{document}